%% file: main.tex
\documentclass[submission]{eptcs}
\usepackage{etex}
\providecommand{\event}{ICE 2013}

\def\finex{{\unskip\nobreak\hfil
\penalty50\hskip1em\null\nobreak\hfil$\diamond$
\parfillskip=0pt\finalhyphendemerits=0\endgraf}}

\newcounter{mmyexample} 

\newcounter{myproposition} 

\newcounter{mytheorem} 
\usepackage{graphicx}
\usepackage{amsmath}
\usepackage{amssymb}
\usepackage{pdfsync}
\usepackage[usenames,dvipsnames]{pstricks}
\usepackage{epsfig}
\usepackage{pst-grad} % For gradients
\usepackage{pst-tree} % For axes
\usepackage[arrow,matrix,frame,curve]{xy}%\CompileMatrices
\usepackage{color,graphics,xcolor}
\usepackage{tikz}
\usepackage{ulem}%\normalem

\title{On Recovering from Run-time Misbehaviour in ADR\thanks{This work has been supported by FP7-PEOPLE-2011-IRSES MEALS}}

\author{Kyriakos Poyias \qquad \qquad Emilio Tuosto
  \institute{Department of Computer Science, University of Leicester,
    UK} \email{kyriakospoyias@gmail.com \qquad \qquad \quad emilio@le.ac.uk}
}

\def\titlerunning{On Recovering from Run-time Misbehaviour in ADR}
\def\authorrunning{K. Poyias, E. Tuosto}

\input{commands}

\begin{document}
\input{tikzmacros}
\maketitle

\begin{abstract}
  We propose a monitoring mechanism for recording the evolution of 
  systems after certain computations, maintaining the history 
  in a tree-like structure.
  Technically, we develop the monitoring mechanism in a variant of ADR
  (after Architectural Design Rewriting), a rule-based formal framework
  for modelling the evolution of architectures of systems.
    
  The hierarchical nature of ADR allows us to take full advantage of the
  tree-like structure of the monitoring mechanism. We exploit
  this mechanism to formally define new rewriting mechanisms for
  ADR reconfiguration rules.
  Also, by monitoring the evolution we propose a way of identifying which part
  of a system has been affected when unexpected run-time behaviours emerge.
  Moreover, we propose a methodology to suggest reconfigurations that
  could potentially lead the system in a non-erroneous state.
\end{abstract}

% Main part of your article
% =========================
\section{Introduction}
\label{sec:intro}
\input{introduction}

%\section{Prerequisites}
%\label{prerequisites}

%\subsection{A Variant of ADR}
\section{A Variant of ADR}
\label{sec:adr}\label{sec:dbc}\label{sec:wpc}
\input{adr}

%\subsection{Design by Contract for ADR}
\input{dbc}

\section{Tracking ADR Architectural Reconfigurations}
\label{sec:tracking}
\input{tracking}

\section{A new rewriting mechanism}
\label{sec:newre}
\input{rewrite}

\section{Recovering invalid configurations}
\label{sec:methodology}
\input{methodology}

\section{Conclusion}
\label{sec:conc}
\input{future}

% Bibliography with BibTeX
% ========================
\bibliographystyle{eptcs}
\bibliography{mybibliography}

\end{document}

%% file: commands.tex
\newcommand{\inter}{\delta}

\newcommand{\addTree}{\mathsf{addTree}}
\newcommand{\child}[2]{{#2}[{#1}]}
\newcommand{\ptm}[2]{{#1} \bowtie {#2}}
\newcommand{\getvartree}[3]{{#1} \ \sqsubset_{#3}\  {#2}}
\newcommand{\talg}{\mathtt{Term}_{\Sigma,{\mathcal X}}}
\newcommand{\troot}{\diamond}
\newcommand{\defref}[1]{Definition~\ref{#1}}
\renewcommand{\deg}[1]{\mathrm{deg}\ {\lvar #1}}
\newcommand{\figref}[1]{Fig.~\ref{#1}}
\newcommand{\secref}[1]{\S~\ref{#1}}

\newcommand{\dom}[1]{\text{dom}\ {#1}}

\newcommand{\hidden}[1]{}
\newcommand{\lvar}[1]{{\mathsf{#1}}}

\newenvironment{prodenv}{\begin{minipage}[c]{6cm}\tiny}{\end{minipage}}
\newenvironment{prodenv2}{\begin{minipage}[c]{5cm}\tiny}{\end{minipage}}

\newcommand{\wpre}[4]{\mathcal{W}^{#1}_{#2}(#3,#4)}

  \definecolor{shadecolor}{rgb}{1,0.99,0.9}

  \definecolor{greenshade}{RGB}{193,255,193}

\newcommand{\curlybraces}[1]{\{ #1 \}}
\newcommand{\anglebraces}[1]{\langle #1 \rangle}
\newcommand{\undef}{\uparrow}

\newcommand{\envempty}{\mathbf{0}}

\newcommand{\treeEnv}{\mathcal{T}}
\newcommand{\tenv}[1]{\treeEnv^{(#1)}}
\newcommand{\treeEnvNot}[3]{\treeEnv({#1}) \stackrel{as}{=} {#2}\cdot{#3}}
\newcommand{\cref}{cf.}
\newtheorem{mydef}{Definition}
\newtheorem{myex}{Example}

\newcommand{\repmap}[1]{\theta_{#1}}
\newcommand{\nodes}[1]{V_{#1}}
\newcommand{\edges}[1]{E_{#1}}
\newcommand{\tentacle}[1]{t_{#1}}
\newcommand{\fv}[1]{\mathrm{fv}{(#1)}}

\newcommand{\true}{\top}
\newcommand{\false}{\bot}

\newcommand{\tg}{\Gamma}
\newcommand{\varset}{\mathsf{V}}
\newcommand{\tnodeset}{\mathsf{N}}
\newcommand{\pset}{\mathcal{P}}

\newcommand{\nodeset}{\mathfrak{N}}
\newcommand{\edgeset}{\mathfrak{E}}
\renewcommand{\L}{\mathcal{L}}
\newcommand{\typing}[1]{\tau_{#1}}
\newcommand{\subs}[2]{[#1 \mapsto #2 ]}
\newcommand{\tsubs}[2]{\subs{\tilde{#1}}{\tilde{#2}}}
\newcommand{\mmdef}{\mbox{$\;\stackrel{\mathrm{def}}{=}\;$}}

\newcommand{\inodes}[1]{{#1}^\circ}
\newcommand{\st}{\ \big|\ }

\newcommand{\set}[1]{{\curlybraces{{#1}}}}
\newcommand{\tentfun}[3]{{#1} : {#2} \mapsto ({#3})}
\newcommand{\gnoL}{G\backslash \sigma(E_L)}

\newcommand{\conf}[1]{\langle {#1} \rangle}
\newcommand{\no}[1]{\mathtt{noEdge}\conf{#1}}

\newcommand{\after}[2]{{#1} \circ {#2}}

\newcommand{\prd}[3]{\set{#1}\ {#2} \ \set{#3}}

\newcommand{\accessport}{\circ}
\newcommand{\chainport}{\bullet}
\newcommand{\tyedge}[1]{*+[F-,]{{#1}}}
\newcommand{\tedge}[1]{*+[F]{{#1}}}
\newcommand{\ntedge}[1]{*+[F=]{{#1}}}

\newcommand{\ntaritythreeT}[4]{\ntedge{{#1}} \ar[{#2}] \ar@{-}[{#3}] \ar@{-}[{#4}]}
\newcommand{\ntaritythree}[5]{\ntedge{{#1}} \ar@(r,l)[{#2}] \ar@{-}@(d,{#3})[{#4}] \ar@{-}@(l,r)[{#5}]}
\newcommand{\ntaritytwo}[3]{\ntedge{{#1}} \ar[{#2}] \ar@{-}[{#3}]}
\newcommand{\ntarityone}[2]{\ntedge{{#1}} \ar[{#2}]}
\newcommand{\taritythree}[4]{\tedge{{#1}} \ar@(r,l)[{#2}] \ar@{-}@(d,u)[{#3}] \ar@{-}@(l,r)[{#4}]}
\newcommand{\taritytwo}[3]{\tedge{{#1}} \ar@(r,l)[{#2}] \ar@{-}@(l,r)[{#3}]}
\newcommand{\tarityone}[2]{\tedge{{#1}} \ar[{#2}]}

\newcommand{\client}{\texttt{C}}
\newcommand{\bank}{\texttt{B}}
\newcommand{\bankobj}[1]{{\mathit{#1}:\bank}}

\newcommand{\bookF}{\texttt{BF}}

\newcommand{\findF}{\texttt{FF}}
\newcommand{\findFobj}[1]{{\mathit{#1}:\findF}}
\newcommand{\flights}{\texttt{Fls}}
\newcommand{\flightsobj}[1]{{\mathit{#1}:\flights}}
\newcommand{\flight}{\texttt{Fl}}
\newcommand{\flightobj}[1]{{\mathit{#1}:\flight}}
\newcommand{\pay}{\texttt{P}}
\newcommand{\payobj}[1]{{\mathit{#1}:\pay}}
\newcommand{\fpay}{\texttt{PF}}

\newcommand{\bookflightprod}{\mathtt{bookFlight}}
\newcommand{\payprod}{\mathtt{pay}}

\newcommand{\browseflightsprod}{\mathtt{browseFlights}}

\newcommand{\chainprod}{\bookflightprod}

\newcommand{\inv}{\phi_\text{inv}}

%% file: tikzmacros.tex
\usetikzlibrary{arrows,shapes,backgrounds,positioning}%,fadings,patterns,decorations.pathreplacing}
\tikzstyle{morphism} = [-triangle 90] \tikzstyle{map} = [-open
triangle 45] \tikzstyle{hookstyle} = [left hook->]

%% file: introduction.tex
We present a technical development of the \emph{Architectural Design
Rewriting} model (ADR)~\cite{bllmt08} that combines the features of
ADR described in~\cite{bllmt08,our:ice2012}.
We take the motivation of our work from the problems arising in modern
software development.
Software systems are no longer static and become more and more dynamic;
because of their very interactive nature, such systems are starting
to be studied under new angles~\cite{gsw06}.
For instance, software needs to adapt to the (often unpredictable)
changes of the (virtual and physical) environment it operates in.
The term \emph{autonomic computing} has been coined to mark such
systems~\cite{hm08}, which present new degrees of complexity since
they require high levels of flexibility and
adaptiveness~\cite{jlny04}.

Such complexity calls for rigorous methods at very early stages of
software development.
Architectural Description Languages (ADLs) used to design such systems
have to be able to guarantee software quality and correctness by 
being flexible to adapt from their initial designs, and also be able 
to predict the possible problems that could arise during the execution
of such systems.
Formal approaches aim to devise robust engineering practices to form
reliable software products to mitigate the issues described above.
Arguably, those approaches focus on software behaviour; correctness
and efficiency of software play in fact a crucial role not only in
critical systems but also in daily-life applications.
In the design phase, semi-formal methods are typically adopted; as an
example, the use of modelling languages is combined with design
patterns to devise a model that can be checked.
This approach may involve formal techniques (e.g., type or model
checking) to guarantee properties of applications while non-formal
techniques (or tools not supported by formal approaches) are typically
used to tackle architectural design aspects.
Our research agenda envisages the combination of those approaches with
techniques to address the issues above at the design level.
We believe that a rigorous treatment at the design level would allow
to identify and solve many problems that are currently tackled only by
inspecting or testing code.

We describe a formal framework that is able to tackle the
architectural/structural aspects of the design and allow designers to
identify and address problems at a higher level.
In~\cite{our:ice2012} we advocate a design-by-contract (DbC approach)
for ADLs that allows the specification of \emph{contracts} that
architectures have to abide by.
On these grounds, in~\cite{our:ice2012} we propose a methodology that
is able to compute structural ``rearrangements'' of systems' architecture
to  adapt themselves when an erroneous state is reached.
Technically, the DbC approach in~\cite{our:ice2012} is developed by
extending ADR~\cite{bllmt08} with \emph{asserted production rules},
that is rules for architectural transformations equipped with logical
\emph{pre-} and \emph{post-conditions}.
In this way, asserted productions become contracts that guarantee the
\emph{architectural style} when they are applied.
The concept of architectural style is crucial in software
architectures~\cite{tmd09}.
In ADR, the architectural style of a system is formalised in terms of
\emph{productions rules}, namely rules that can be used to generate
specific configurations of the architectural elements.
As surveyed in \S~\ref{sec:adr}, ADR models architectures as
\emph{(hyper)graphs} that is a set of \emph{(hyper)edges} sharing some
nodes; respectively, edges represent architectural elements (at some
level of abstraction) while nodes represent components' interfaces.
Also, ADR production rules take the form $p : L \to R$ where $L$ is an
edge and $R$ an (hyper)graph; rule $p$ is meant to replace $L$ with
$R$ within a given graph.
The main contribution in~\cite{our:ice2012} is an algorithm that
computes a weakest pre-condition $\psi$ out of a post-condition
$\varphi$ and a production rule.
We prove a theorem that guarantees that the application of the rule to
a configuration satisfying $\psi$ yields a configuration satisfying
the post-condition $\varphi$.
This algorithm can be used to compute a reconfiguration if the current
configuration violates the invariant.
Roughly, in~\cite{our:ice2012} we envisage architectural styles
according to the equation:
\begin{eqnarray}\label{eq:style}
  \text{architectural style} & = & \text{production rules} + 
  \text{invariants}
\end{eqnarray}
where an invariant is the property the designer requires of the
application.

\paragraph{A summary of our contributions. } The main contributions of
this paper can be summarised as an extension and a refinement of
the methodology in~\cite{our:ice2012}.

The extension consists in the adaptation of the methodology to
encompass \emph{reconfiguration mechanism} of ADR.
As a matter of fact, ADR features complex reconfigurations that
cannot be captured by production rules.
Such complex reconfigurations can be envisaged as a model of run-time
evolution of systems that describe what complex rearrangements could
happen during execution.
This is technically done by specifying term rewriting rules in
an algebra where terms are interpreted as proofs of  the style
of graphs.
Here we broaden the applicability of the methodology
in~\cite{our:ice2012} to a more general setting that allows the
iteration of the methodology in~\cite{our:ice2012} when reconfigurations
violate the style.
A limitation of the methodology in~\cite{our:ice2012} was due to the
fact that style violations could be fixed only considering
``top-down'' application of productions.
In this paper we take into account also violations of the style due to
complex reconfigurations.
To achieve this we have to identify the ``positions'' in the system
where violations occur.
This allows us to apply our methodology only to the parts of the system
affected by the ill reconfiguration.
Also, we propose here a systematic reiteration of the methodology when
an immediate way to recover the style cannot be found.
This yields a more general and efficient framework.

Intuitively, the equation~\eqref{eq:style} now becomes
\[
  \text{architectural style} \ = \ \text{production rules} +
  \text{reconfiguration rules} + 
  \text{invariants}
\]
This generalisation is possible due to the introduction of 
 a monitoring approach that fully exploits the features
of ADR.

The refinement we propose here regards the rewriting mechanism of ADR
and, more importantly, its underlying monitoring capability.
Indeed, as observed in~\cite{bllmt08}, a distinctive aspect of ADR is
that it features the canonical view of software architectures in terms
of connected architectural elements as well as a hierarchical view of
software architectures that is paramount in the design phase.
Although~\cite{bllmt08} advocates the use of the architectural view as
a useful mechanism to be exploited in complex reconfigurations, no
actual formalisation has been provided on how this could be achieved
and the parsing features of ADR had been only sketched
in~\cite{bllmt08}.
More precisely, we start by proposing minor changes to the original
rewriting mechanism of ADR that distinguishes edge as terminal and
non-terminal at the type level and allows only non-terminal edges to
be rewritten (for each edge type, either all the edges of that type
can be rewritten or none of them).
Our generalisation eliminates such distinction; an edge can be rewritten
if it is marked as ``replaceable'' in the graph.
Therefore, we allow edges of the same type to be rewritten or not
depending on how they are marked in the graph.
(We note that the original ADR rewriting mechanism can still be
obtained: if one decides that an edge type is non-terminal, then all
edges of that type have to be replaceable while for types of terminal
edges, all edges have to be marked as non-replaceable.)
Besides some simplification in the technical presentation of ADR (which
is now more uniform), such generalisation brings in extra flexibility.
In fact, a replaceable edge can be refined by introducing new versions
of a rule that differs only for the ``replaceability'' of some edges.

In addition we introduce a monitoring mechanism that is also exploited
to define an efficient parsing of ADR graphs.
Our monitoring mechanism keeps track of the application of
reconfiguration rules and uses such information when the graph has to
be parsed (to identify the part that violates a style).
The application of a reconfiguration rule affects such information
that need to be updated accordingly.

Summing up, one can enforce the architectural style of the system in
presence of complex reconfiguration that may violate the style; this
can be achieved by
\begin{enumerate}
\item defining  a monitoring mechanism,
\item repeatedly adapting the methodology in~\cite{our:ice2012}
  exploiting the parsing features defined here.
\end{enumerate}

\paragraph{Structure of the paper.} \secref{sec:adr} overviews ADR and
introduces its new variant as well as it summarises the results
in~\cite{our:ice2012}.
\secref{sec:tracking} defines our monitoring approach.
\secref{sec:newre} gives the new rewriting mechanism hinging on
our monitoring approach.
\secref{sec:methodology} gives the refinement of our methodology.
\secref{sec:conc} draws some conclusions.

%%% Local Variables: 
%%% mode: latex
%%% TeX-master: "main"
%%% End: 

%% file: adr.tex
In the following, $\nodeset$ and $\edgeset$ are two countably infinite
and disjoint sets (of nodes and edges respectively), $X^* \mmdef
\set{(x_1,\ldots,x_n) \st x_1,\ldots,x_n \in X}$ is the set of finite
lists on a set $X$, and $\tilde x$ ranges over $X^*$.
Also, abusing notation, we sometimes use $\tilde x$ to
indicate its underlying set of elements.

\begin{mydef}[(Hyper)graphs and morphisms~\cite{bllmt08}]\label{graphsDef}\label{morphismDef}
  A \emph{(hyper)graph} is a tuple $G = \anglebraces{V, E, t}$ where
  $V \subseteq \nodeset$ and $E \subseteq \edgeset$ are finite and $t
  : E \to V^*$ is the \emph{tentacle function} connecting edges $e \in
  E$ to a list of nodes; the \emph{arity} of $e$ is
  the length of $\tentacle G(e)$.
  It is convenient to write $e(\tilde u) \in G$ for $e \in \edges G$,
  $\tentacle G(e) = \tilde u \subseteq \nodes G$; also, given a graph
  $G$, $\nodes G$, $\edges G$, and $\tentacle G$ respectively denote
  the nodes, the edges, and the tentacle function of $G$.
  
  Given two graphs $G$ and $H$, a \emph{morphism from $G$ to $H$}
  is a pair of functions $\anglebraces{ \sigma_V:\nodes G \to \nodes
    H, \sigma_E :\edges G \to \edges H}$ s.t. $\sigma_V$ and
  $\sigma_E$ preserve the tentacle functions, i.e.
  $\after{\sigma_V^*}{\tentacle G} = \after{\tentacle H}{\sigma_E}$,
  where $\sigma_V^*$ is the homomorphic extension of $\sigma_V$ to
  $\nodes G^*$.
\end{mydef}

In ADR, graphs are typed over a fixed type graph via typing morphisms.
As usual an ADR graph \emph{$G$ is typed over a type graph $\tg$
  through $\typing G$} if $\typing G$ is a morphism from $G$ to $\tg$.
\begin{mydef}[Typed graphs]\label{def:typed}
  Let $\tg$ be a type graph.
  An ADR graph \emph{$G$ is a (hyper)graph typed over $\tg$
    through $\typing G$} if
  $\typing G$ is a morphism from $G$ to $\tg$.
\end{mydef}
%
%This is reminiscent of string grammars where terminal symbols
%correspond to terminal edges and non-terminal symbols to non-terminal
%edges.

%\kP{Add new Example with no Terminal/non-terminal edges}{Fix:} 
\newcommand{\ff}{\mathit{ff}}
\newcommand{\fl}{\mathit{fl}}
\begin{myex}\label{ex:tgg}
  Take the type graph $\tg = \anglebraces{V, E, t}$ where $V =
  \set{\chainport,\accessport} \subseteq \nodeset$, $E =
  \set{\client, \bank, \findF, \flights, \flight, \bookF, \pay, \fpay}
  \subseteq \edgeset$, and $\tentfun t \client {\chainport}$, $\tentfun
  t \bank {\chainport,\accessport}$, and $\tentfun t e {\chainport ,
    \chainport} $ for each $e \in E \setminus \set{\bank , \client}$.
  \\The graph $G = \anglebraces{\set{u_1, u_2, u_3,
      u_4},\set{\ff,\fl_1,\fl_2}, t'}$ where $t'$ is defined as
  $\tentfun{t'}{\ff}{u_2,u_1}$, $\tentfun{t'}{\fl_1}{u_3,u_2}$, and
  $\tentfun{t'}{\fl_2}{u_4, u_2}$ can be typed on $\tg$ by $\typing G$
  mapping all the nodes to $\bullet$, $\fl_1$ and $\fl_2$ to $\flights$,
  and $\ff$ to $\findF$.
\end{myex}

Hereafter, we fix a typed graph $\tg$ and tacitly assume that all
graphs $G$ are typed over $\tg$ via a morphism $\typing G$.
Intuitively, $\tg$ yields the \emph{vocabulary} of the architectural
elements to be used in the designs; moreover, $\tg$ specifies how
these elements can be connected together (e.g., as in
Example~\ref{ex:tgg}).

For technical reasons we introduce a slight variant of ADR;
instead of considering edges as \emph{non-terminal} 
and \emph{terminal} edges, the new version of ADR allow more
liberal rewriting mechanism by marking in a graph the edges
that can be replaced.
Technically, this is obtained by considering pairs $\conf{G,\theta}$
where $G$ is a graph and $\theta : \edges G \to \set{0,1}$ is the
\emph{replaceability map}; an edge $e \in \edges G$ is
\emph{replaceable} iff $\theta(e) = 1$.
Abusing notation we will implicitly assume that any graph $G$ is equipped
with a replaceablility map which we will denote by $\repmap G$. 

Type and typed graphs have a convenient visual notation.
Nodes are circles and edges are drawn as (labelled) boxes; tentacles
are depicted as lines connecting boxes to circles; conventionally,
directed tentacles indicate the first node attached to the edge and
the others are taken clockwise.
The boxes of edges of type graphs are shaded, while the edges in a
graph are either single- and double-lined boxes; the former represent
non-replaceable edges while the latter represent replaceable ones.
The visual notation for typed graphs include the graph and its typing
morphism.
Nodes are paired with their types while an edge label $e : e'$
represents the fact that the typing morphism maps the edge $e$ of the
graph to the edge $e'$ of the type graph.
%
%\kP{Add new Examples}{Update:} 
\begin{myex}\label{ex:tgandg}
  In the visual notation described above, the type graph $\tg$ and the
  graph $G$ of Example~\ref{ex:tgg} can be respectively drawn as
  {\small\[
  \begin{array}{c@{\hspace{2cm}}c}
    \xymatrix@C=1cm@R=.7cm{
      \tyedge{\client} \ar@(r,l)[r]
      & {\chainport}
      & \tyedge{e} \ar@(d,dr)[l] \ar@{-}@(l,r)[l]
      \\
      \accessport & \tyedge \bank \ar[u] \ar@{-}[l]
    }
    &
    \xymatrix@C=1cm@R=.7cm{
      & & &
      \ntedge{\flightsobj{\fl_1}} \ar@(r,l)[r] \ar@{-}@(l,ur)[dl]
      &\stackrel{u_3}{\chainport}\\
      \stackrel{u_1}{\chainport} &
      \tedge{\findFobj{\ff}} \ar@(r,l)[r] \ar@{-}@(l,r)[l]
      & \stackrel{u_2}{\chainport} &
      \ntedge{\flightsobj{\fl_2}} \ar@(r,l)[r] \ar@{-}@(l,r)[l]
      & \stackrel{u_4}{\chainport} &
    }
  \end{array}\]}
where, to simplify the type graph, we use $e \in E
\setminus\set{\bank , \client}$ (instead of drawing an edge for each
edge of $\Gamma$ with arity two).
\end{myex}

\begin{mydef}[Typed Graph morphisms]\label{tgDef}
  A \emph{morphism} between $\tg$-typed graphs $f: G_1 \to G_2$ is a
  \emph{typed graph morphism} if it preserves the typing, i.e. such
  that $\typing{G_1} = \typing{G_2} \circ f$.
\end{mydef}
Note that replaceablility maps are not considered in
Definition~\ref{def:typed}.

\begin{mydef}[Productions]\label{dpDef}
  A \emph{(design) production} $p$ is a tuple $\anglebraces{ L,R,
    i:\nodes L \to \nodes R}$ where $L$ is a graph consisting of a
  single repleaceble edge attached to distinct nodes and $R$ is a
  graph; the nodes in $Im(i)$ (the image of $i$) are called
  \emph{interface nodes}.
\end{mydef}

Design productions can be thought of as rewriting rules that, when
applied to a graph $G$, replace 
a replaceable (hyper)edge of $G$ matching $L$
 with a fresh copy of
$R$ (we remark that
our morphisms are type-preserving).

\begin{myex}\label{ex:prod}
  Take the following graphs:
  \begin{eqnarray*}
  G_L & = & \anglebraces{\{a,b\}, \{fs\}, fs \mapsto (a,b)}
  \\
  G_R & = & \anglebraces{\{u_1,u_2,u\}, \{fls,pa\}, t_R:\begin{cases}fls \mapsto
    (u,u_2) \\ pa \mapsto (u_1,u) \end{cases}}
  \end{eqnarray*}
  with $fs$  of type type $\flights$, $fls$ of type $\flight$, and
  $pa$ of type $\pay$.
  (Note that $G_L$ is a single-edge graph.)

  The production $\bookflightprod = \anglebraces{G_L, G_R, i_\bookflightprod}$
  has $G_L$ and $G_R$ as left-hand side (LHS) and right-hand side
  respectively; the interface of $\bookflightprod$ is given by the map
  defined as follows:
  \[
  i_\bookflightprod: \
  \begin{cases}
    \ a \mapsto u_1
    \\ 
    \ b \mapsto u_2
  \end{cases}
  \]
  namely, $a$ (resp. $b$) corresponds to the first (resp. second) node
  of $fs$.
\end{myex}

Like ADR graphs, productions have an appealing visual representation
that we illustrate in the next example that depicts the production of Example
\ref{ex:prod}.
\begin{myex}\label{ex:chainProds}
  The graphical representation below corresponds to the production $\bookflightprod$ in Example~\ref{ex:prod}.
  \[\begin{minipage}{7.5cm}{$ \qquad \bookflightprod
      \\
      \xymatrix@C=.3cm@R=.3cm{ & \flightsobj{fs}
        \\
        \stackrel{b}{\chainport} \ar@{.}[r] &
        \stackrel{u_2}{\chainport} & \ntaritytwo{\ \ \flightobj{fls} \ \
        }{r}{l} & \stackrel{u}{\chainport} & \ntaritytwo{\ \
          \payobj{pa}\ \ }{r}{l}& \stackrel{u_1}{\chainport} &
        \stackrel{a}{\chainport} \ar@{.}[l] \save "1,2"."2,6"*+[F.] \frm{} \restore
      } $}
    \end{minipage}
  \]
  The dotted square and the dotted lines represent the LHS and the map
  $i_\bookflightprod$; the name and type of the edge of the LHS is in the
  top-left corner of the dotted box and the name of the production is
  given on the top of the dotted square.
  The RHS of $\bookflightprod$ is depicted inside the dotted box
\end{myex}
%%%%%%%%%%%%%%%%%%%%%%%%%%%%%%%%%%%%%%%%%%%%%%%%%%%%%%%%%%%%%%%%%%%%%%%%%%%%%%%%%%%%%%%%%
The next example illustrates how productions are applied to graphs;
the details will be given in \secref{sec:dbc} for \emph{asserted
productions}, which encompass ADR productions.

\begin{myex}\label{ex:chainApplic}
  Consider the production $\chainprod$ of Example~\ref{ex:chainProds}.
  Below, the unique edge of type $\flights$ is replaced by an instance
  of the RHS of $\bookflightprod$.
  \[\begin{array}{c@{\hspace{.5cm}}c}
      \def\g#1{\save [].[dr]!C="g#1"\frm{}\restore}%
      \xymatrix@C=.5cm@R=.3cm{
      % g1r1
      \tedge{\findFobj{ff}}
      \ar@(r,l)[r] \ar@(d,u)@{-}[dd]
      &\stackrel{u_1}{\chainport}&&
      % g2r1
      \g2&
      \tedge{\findFobj{ff}}
      \ar@(r,l)[rr] \ar@(d,u)@{-}[dd]
      &&\stackrel{u_1}{\chainport}
      &   \\&&&&\\
      % g1r2
      \stackrel{u}{\chainport}
      &\ntedge{\flightsobj{fls}}
      \ar@(r,r)[uu] \ar@(u,d)@{-}[uu]
      &&
      % g2r2
      & \stackrel{u}{\chainport}
      & \ntedge{\flightobj{f}}
      \ar@(r,l)[r] \ar@(u,dl)@{-}[uur]
      &\stackrel{u_2}{\chainport}
      & \ntedge{\payobj{p}}
      \ar@(r,r)[uul] \ar@(l,r)@{-}[l]
      \ar @{=>} "2,3" ;"2,4" ^-{\bookflightprod}
      % \ar @{-->} "2,1";"g2" 
    }
    \end{array}\]
  Note that the rest of the graph (consisting only of the edge $\ff$)
  including the interface nodes is left unchanged while a fresh node
  $u_2$ is created.
\end{myex}
%%%%%%%%%%%%%%%%%%%%%%%%%%%%%5

%%% Local Variables: 
%%% mode: latex
%%% TeX-master: "main"
%%% End: 

%% file: dbc.tex
We overview the Design by Contract (DbC) approach for ADR introduced
in~\cite{our:ice2012}.
Note that the variant of ADR given in \secref{sec:adr} generalises the
rewriting mechanism originally defined in~\cite{bllmt08}, therefore
the results in~\cite{our:ice2012} can be easily adapted to the variant
of ADR presented here.

Properties of graphs are expressed in a simple logic tailored for ADR.
In the following we let $D, D', \ldots$ range over edges of $\tg$.
\begin{mydef}[ADR logic~\cite{our:ice2012}]\label{def:logic}
  Let $\varset$ be a countably infinite set of variables for nodes
  (ranged over by $\lvar x, \lvar y, \lvar z, \ldots$).
  The set $\L$ of \emph{(graph) formulae} for ADR is given by the
  following grammar:
  \[
  \psi, \varphi \quad ::=  \quad \lvar x = \lvar y
  \quad | \quad      \true
  \quad | \quad      \neg \varphi
  \quad | \quad      \varphi_1 \land \varphi_2   
  \quad | \quad      \forall D(\tilde{\lvar x}).\varphi
  \]
  In formulae of the form $\forall D(\tilde{\lvar x}).\varphi$, the
  occurrences of $\lvar y \in \tilde{\lvar x}$ in $\varphi$ are
  \emph{bound}, $\tilde{\lvar x}$ has the length of the arity of $D$
  and $\tilde{\lvar x}$ are pairwise distinct.
\end{mydef}
Basically, $\L$ is a propositional logic to predicate on
(in)equalities of nodes and it is parametrised with respect to the
type graph $\tg$ used in quantification.
Variables not in the scope of a quantifier are free and the set $\fv
\varphi$ of \emph{free variables} of $\varphi \in \L$ is defined
accordingly; also, we abbreviate $\lvar x_1 = \lvar x_2 \land \ldots
\land \lvar x_{n-1} = \lvar x_n$ with $\lvar x_1 = \lvar x_2 = \ldots
= \lvar x_{n-1} = \lvar x_n$ and we define $\bot$ as $\neg \top$,
$\lvar x \neq \lvar y$ as $\neg (\lvar x = \lvar y)$, $\varphi \vee
\psi$ as $\neg ( \neg \varphi \land \neg \psi)$,  $\varphi \to \psi$
as $\neg \varphi \vee \psi$ , and $\exists D(\tilde{\lvar x}).\varphi$
as $\neg \forall D(\tilde{\lvar x}).\neg \varphi$.
The models of $\L$ are ADR graphs together with an interpretation of
the free variables of formulae.
\begin{mydef}[Satisfaction relation]\label{def:entailment}
  A graph $G$ \emph{satisfies $\varphi \in \L$ under the assignment $h
    : \varset \to \nodes G$} (in symbols $G \models_h \varphi$) iff
  \[\begin{array}{lcll}
    \varphi \equiv \true,
    & & & or 
    \\
    \varphi \equiv \lvar x = \lvar y
    & \text{and} &
    h(\lvar x) = h(\lvar y),
    & or
    \\
    \varphi \equiv \neg \varphi'
    & \text{and} &
    G \nvDash_h \varphi' ,
    & or
    \\
    \varphi \equiv \varphi_1 \land \varphi_2 
    & \text{and} &
    G \models_h \varphi_1 \text{ and } G \models_h \varphi_2 ,
    & or
    \\
    \varphi \equiv \forall D(\tilde{\lvar x}).\varphi
    & \text{and} &
    G \models_{h \tsubs {\lvar x} u} \varphi
    \; \text{ for any } d(\tilde u) \in G &  \text{s.t. } \tau_G(d)= D
  \end{array}\]
\end{mydef}
Note that in the last clause of Definition~\ref{def:entailment}, each
bound variable in $\tilde{\lvar x}$ is instantiated with a node.
It is easy to prove that we can restrict to finite mappings that only assign the
free variables of formulae.
Namely, for each $h,h' : \varset \to \nodes G$, if $h|_{\fv \varphi} =
h'|_{\fv \varphi}$ then $G \models_h \varphi$ iff $G \models_{h'}
\varphi$.
We write $G \models \varphi$ when $\fv \varphi = \emptyset$.

\begin{myex}\label{ex:test}\label{ex:interesting}
  Consider 
  the formulae
  \begin{eqnarray}\label{eq:notexists}
    \no D & \mmdef & \forall D(\tilde{\lvar x}).\false
    \\
    \phi_{\text{ex}} & \mmdef & \forall D(\lvar x, \lvar y).\exists D'(\lvar z). \lvar x = \lvar z
    \label{eq:valid}
  \end{eqnarray}
  Formula~\eqref{eq:notexists} characterises the graphs that do not
  contain edges of a given type while the formula~\eqref{eq:valid}
  describes graphs such that each edge of type $D$ is connected to one
  of type $D'$ on the first tentacle.
  For instance, consider the graphs
  {\scriptsize\[
    \begin{array}{c@{\hspace{1cm}}c}
        G_{valid} =
      \xymatrix@C=.4cm@R=.2cm{
        \stackrel{u_2}{\chainport}
        &\tedge{d_1:D} \ar@(r,l)[r] \ar@{-}@(l,r)[l]
        &\stackrel{u_1}{\chainport}
        &\tedge{d':D'} \ar@(l,r)[l]\\
        \stackrel{u_4}{\chainport} 
        &\tedge{d_2:D} \ar@(r,d)[ur] \ar@{-}@(l,r)[l]
      }
      &
      G_{invalid}  =
      \xymatrix@C=.4cm@R=.2cm{
        \stackrel{u_2}{\chainport}
        &\tedge{d_1:D} \ar@(r,l)[r] \ar@{-}@(l,r)[l]
        &\stackrel{u_1}{\chainport}
        &\tedge{d':D'} \ar@(l,r)[l]\\
        \stackrel{u_4}{\chainport} 
        &\tedge{d_2:D} \ar@(r,l)[r] \ar@{-}@(l,r)[l]
        &\stackrel{u_3}{\chainport} 
      }
    \end{array}\]}
  then $G_{valid}$ satisfies $\phi_{\text{ex}}$ whereas $G_{invalid}$
  does not, because $d_2$ is not connected to any edge of type $D'$.
\end{myex}

Fix an ADR production  $p = \anglebraces{L,R,i}$.
Our notion of contracts hinges on \emph{asserted productions}, namely
ADR productions decorated with pre- and post-conditions expressed in
the logic $\L$.
Given $\psi, \varphi \in \L$ and two assignments $h, h': \varset \to
\nodeset$, an \emph{asserted production} is an expression of the form
\begin{equation} \label{def:assertedprod}
  \prd{\psi , h} p {\varphi , h'}
  \qquad \text{where } \quad
  h(\fv \psi) \subseteq \nodes L \text{ and } h'(\fv \varphi) \subseteq \nodes R
\end{equation}
An asserted production generalises ADR productions and it intuitively
requires that if $p$ is applied to a graph $G$ that satisfies $\psi$
then the resulting graph is expected to satisfy $\varphi$.
The assignments $h$ and $h'$ in~\eqref{def:assertedprod} allow pre-
and post-conditions to predicate on nodes occurring in the LHS or the
RHS of $p$.

Operationally, an asserted production $\pi$ can be applied to a graph
$G$ by replacing an ``instance'' of the LHS in $G$ (identified by an
\emph{matching homomorphism}) with a new instance of the RHS and
connecting the interface nodes accordingly, provided that $G$
satisfies the precondition of $\pi$ (under the matching homomorphism).
For the variant of ADR proposed in \secref{sec:adr} we just have to
impose the condition that the homomorphic image of the LHS of $\pi$
has to be a replaceable edge.
This is schematically illustrated in Figure~\ref{fig:dpT}
(cf.~\cite{our:ice2012}) and demonstrated in
Examples~\ref{ex:applicability} and~\ref{ex:application}.
\begin{figure}[t]\centering
  $\xymatrix@R1.5em@C3em{
    \psi \ar[r]^h & L\ar[r]^i\ar[dd]^{\sigma} & R\ar@{=}[d]^\iota & \ar[l]_{h'}\varphi
    \\
    &   & R'\ar@{^{(}->}[d]
    \\
    & G\ar@{|=}[uul]\ar[r]^\pi & G'
  }$
  %\kP{Generalize diagram a bit}{Note:}
  \caption{\label{fig:dpT}Asserted design productions}
\end{figure}
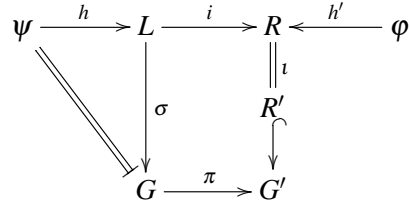
\begin{myex}\label{ex:applicability}
  Consider the production $\bookflightprod$ given in
  Example~\ref{ex:chainProds} and the asserted production
  \[
  \pi \ \mmdef \ \prd{\psi,\emptyset} \bookflightprod {\true,\emptyset}
  \qquad \text{where} \qquad
  \psi \ \mmdef \ \forall \flights(\lvar x, \lvar y).\lvar x \neq \lvar y
  \]
  Then, $\pi$ cannot be applied to the leftmost graph $G$ in the
  rewriting of Example~\ref{ex:chainApplic} because $G \not\models
  \psi$ (under the unique morphism $\sigma$ from $L$ to $G$).
  In fact, $\lvar x$ and $\lvar y$ are mapped to the same node
  $u_1$ of $G$.
\end{myex}
\begin{myex}\label{ex:application}
  The rewriting below is obtained by applying $\pi$ in
  Example~\ref{ex:applicability}.
  \[
    \def\g#1{\save
      [].[ddr]!C="g#1"\frm{}\restore}%
    \xymatrix@C=.35cm@R=.1cm{
      % g1r1
      \tedge{\findFobj{\ff}}
      \ar@(r,l)[r] \ar@(d,u)@{-}[dd]
      &\stackrel{u_1}{\chainport}&&&&
      % g2r1
      \g2&
      \tedge{\findFobj{\ff}}
      \ar@(r,l)[r] \ar@(d,u)@{-}[dd]
      &\stackrel{u_1}{\chainport}
      &
      \\&&&&&&&\\
      % g1r2
      \stackrel{u}{\chainport}
      &\ntedge{\flightsobj{fls}}
      \ar@(r,l)[r] \ar@(u,ld)@{-}[uu]
      & \stackrel{u_3}{\chainport}
      &
      % g2r2
      &&& \stackrel{u}{\chainport}
      & \ntedge{\flightobj{f}}
      \ar@(r,l)[r] \ar@(u,dl)@{-}[uu]
      &\stackrel{u_2}{\chainport}
      & \ntedge{\payobj{p}}
      \ar@(r,l)[r] \ar@(l,r)@{-}[l]
      & \stackrel{u_3}{\chainport}
      \ar @{=>} "2,4" ;"2,6" ^-{\bookflightprod}
      % \ar @{-->} "2,1";"g2" 
    }
  \]
  The edge $\mathit{fls}$ is replaced by an isomorphic instance of $R$
  preserving the interface nodes $u_1$ and $u_3$.
\end{myex}
Note that the application of an asserted production generalises the 
hyper-edge replacement mechanism of ADR; in fact, 
$\prd{\true,\emptyset} p {\true,\emptyset}$ applies exactly as 
normal ADR productions.

\newcommand{\theeq}{\lvar x_1 = \lvar x_2}
An asserted production $\pi$ is \emph{valid} when any application of
$\pi$ to a graph satisfying the precondition of $\pi$ yields a graph
satisfying the post condition of $\pi$.
Obviously, not all asserted productions $\prd{\psi,h} p {\varphi,h'}$
are valid (this can be trivially noted by taking $\varphi$ to be
$\false$).
In~\cite{our:ice2012} we define an algorithm\footnote{ For simplicity,
  we ignore the assignments and environments that the algorithm
  in~\cite{our:ice2012} uses to compute weakest preconditions.} $\wpre
{}{h'} p \varphi$ that, given a production $p$ and a post-condition
$\varphi$, returns a weakest pre-condition $\psi$ so that $\prd{\psi,
  h} p {\varphi, h'}$ is valid.

The next example is adapted from~\cite{our:ice2012}.
\begin{myex}\label{ex:wpexample}
  Consider $\varphi \in \L$ and the production $\payprod$ below:
  \[ \begin{array}{c@{\hspace{1cm}}c}
    \varphi \mmdef \forall B(\lvar x, \lvar y). \forall C(\lvar z).\lvar y = \lvar z
    &
    \payprod \mmdef \begin{minipage}{6cm}{$
        \xymatrix@C=.3cm@R=.01cm{
          &{\pay}
          \\
          & & & \stackrel{u}{\accessport}
          & \tedge{\bankobj b} \ar[r] \ar@{-}[l]&
          \stackrel{u_1}{\chainport} && \chainport^v \ar@{.}[ll]  
          \save "1,2"."2,6"*+[F.] \frm{} \restore
        }$}
    \end{minipage}
  \end{array}\]
  We remark that the post-condition $\varphi$ requires that every edge of
  type $B$ is connected (on its second tentacle) to every edge of type
  $C$.
  The computed weakest pre-condition is
  \[
  \wpre {} {\emptyset} \payprod \varphi \; = \; \no C \; \land \; 
  \forall B(\lvar  x, \lvar y). \forall C(\lvar z).\no C \; \land \;
  \forall B(\lvar  x, \lvar y). \forall C(\lvar z).\lvar y = \lvar z
  \]  
  We remark that $\wpre {} {\emptyset} \payprod \varphi$ imposes
  that for the validity of the asserted production it is necessary
  that the graph does not have any edges of type $C$.
  In fact, production $\payprod$ will generate an edge of type $B$ whose
  second tentacle is attached to an internal node $u$ that cannot be
  shared with any edge of type $C$ already appearing in the graph.
\end{myex}

%%% Local Variables: 
%%% mode: latex
%%% TeX-master: "main"
%%% End: 

%% file: tracking.tex
A key aspect of ADR is to envisage systems as ensembles of
\emph{designs}, that is components \emph{with interfaces}.
Designs are supposed to be generated by means of productions
and can be subject to run-time reconfigurations modelled as
\emph{reconfiguration rules}.
The use of productions yields two pivotal ingredients of ADR.
For clarity, we consider only productions as it is just a matter of
technicality to adapt this section to asserted productions.

Firstly, productions implicitly equip designs with a hierarchical
structure that can be formalised as the ``derivation tree'' determining
the design.
In fact, a set of ADR productions induces a multi-sorted algebraic
signature $\Sigma$ where the sorts are the type edges in the type
graph\footnote{In the original ADR presentation, the sorts are just
  the non-terminal edges. In our variant, this can be simplified by
  taking all edges of the type graph as sorts.} and the operations are
the productions themselves, once a total order on the edges in the RHS
of the production is fixed.
Hereafter, we fix such an order\footnote{The chosen order is
  completely arbitrary and does not affect the construction described
  above.} and, given the RHS $R$ of a production, we write $R[j]$ for the
$j$-th edge in $R$.
With this construction, an ADR production becomes an operation with
type
\begin{equation}\label{eq:prodop}
  E_1 \times \ldots \times E_n \to L
\end{equation}
where $E_k$ is the type of the $k$-th edge in the RHS of the production
(according to the chosen order on edges in the RHS)
and $L$ is the type of the edge in the LHS.
In other words, an ADR production like in~\eqref{eq:prodop} can be
envisaged as an operation in some \emph{algebras of designs} that
builds a design $G$ of type $L$ out of designs $G_k$ of type $E_k$
(for $1 \leq k \leq n$).
This corresponds to a ``bottom-up'' development (whereby designs are
assembled out of other components) and, as observed in~\cite{bllmt08},
it parallels the ``top-down'' generation of designs (similar to
context-free grammars) reviewed in~\secref{sec:adr}.
Moreover, one could consider the terms (with sorted variables to model
partial designs) built on $\Sigma$ and adopt the obvious operational
interpretation: $G$ is obtained by replacing the $j$-th edge in the
RHS with $G_j$ (and connecting the interface nodes as appropriate).
The elements of such term algebra correspond to the proof that a given
design can be assigned some type. 

\begin{myex}\label{ex:signature}
  The production $\bookflightprod$ in Example~\ref{ex:chainProds}
  yields the operation
  \[
  \bookflightprod : \flight \times \pay \to \flights
  \]
  assuming that in the chosen order, $f$ is smaller than $p$.
\end{myex}

Secondly, ADR exploits the algebraic view of productions to model
complex architectural \emph{reconfigurations} that cannot be captured
by productions.
In fact, design can evolve for instance when components have to be
removed, added, or assembled in a different way.
Architectural reconfigurations are naturally modelled as transformation
of elements in the $\Sigma$-term algebra with variables $\talg$ (where
$\mathcal X$ is the set of variables).
Formally, this is achieved by defining a term rewriting system on
$\talg$; namely, a reconfiguration rule takes the form
\begin{equation}\label{eq:reconfiguration} t \to t' \end{equation}
where $t, t' \in \talg$ are linear terms (that is each variable occurs
at most once in $t$ and similarly for $t'$) and the variables
occurring in $t'$ also occur in $t$.

\begin{myex}
  Combining the operation in Example~\ref{ex:signature} with the one
  associated to the production in Example~\ref{ex:chainProds} one could
  build the term $\bookflightprod(x, \payprod(y))$ of type $\flights$
  (provided that $x$ is of type $\flight$ and $y$ is of type $B$.
\end{myex}
\newcommand{\bookf}{\mathtt{bookF}}
Below we give an example of simple reconfigurations.
\begin{myex}\label{ex:reconfiguration}
  Consider the following productions:
  \[
  \begin{array}{c}
    \begin{prodenv2}
      $\browseflightsprod$
      \\
      \xymatrix@C=.3cm@R=.2cm{
        &
        \flight
        &
        & \ntedge{\flightobj{f_1}} \ar[r] \ar@{-}@(l,u)[dl]
        &\stackrel{x_3}{\chainport} 
        \\
        \chainport \ar@{.}[rr] & & \stackrel{x_2}{\chainport} 
        & \ntaritytwo{\flightobj{f_2}}{r}{l}
        & \stackrel{x_1}{\chainport}
        & \chainport \ar@{.}[l]
        \save "1,2"."2,5"*+[F.] \frm{} \restore
      }
    \end{prodenv2}
    \hfill
    \begin{prodenv}
      $\bookf$
      \\
      \xymatrix@C=.3cm@R=.2cm{
        &
        \flight
        \\
        \chainport \ar@{.}[r] & \stackrel{x_2}{\chainport} &
        \ntaritytwo{\ \ \flightobj{f} \ \ }{r}{l} &
        \stackrel{x}{\chainport} &
        \ntaritytwo{\ \ \mathtt{c:Client}\ \ }{r}{l}&
        \stackrel{x_1}{\chainport} & \chainport \ar@{.}[l]
        \save "1,2"."2,6"*+[F.] \frm{}
        \restore }
    \end{prodenv}
  \end{array} 
  \]
  We can define the following reconfiguration rule:
  \[ \mathtt{cf} \ : \browseflightsprod ( x , \bookf(y,z) )
  \rightarrow \browseflightsprod ( \bookf(x,z), y )\]
%  \[ \mathtt{cf2} \ : \browseflightsprod (\bookf(y,z) , x ) \rightarrow \browseflightsprod ( \bookf(y,z), \bookf(x,z) )\]
    \[
    \begin{array}{c@{\hspace{.5cm}}c@{\hspace{.2cm}}c@{\hspace{.2cm}}c@{\hspace{.2cm}}c}
      \begin{minipage}{2cm}
        \xymatrix@C=.3cm@R=.3cm{
          &\ntedge{\flightobj{f_1}} \ar[r] \ar@{-}@(l,u)[dl]
          &\stackrel{u_3}{\chainport} 
          \\
          \stackrel{u_2}{\chainport} 
          & \ntaritytwo{\flightobj{f_2}}{r}{l}
          & \stackrel{u_1}{\chainport}
          &\ntaritytwo{\ \ \mathtt{c:Client}\ \ }{r}{l}&
          \stackrel{u_4}{\chainport}
        }
      \end{minipage}
      &\stackrel{\mathtt{cf}}{\Longrightarrow}&
      \begin{minipage}{2cm}
      \xymatrix@C=.3cm@R=.3cm{
        &\ntedge{\flightobj{f_1}} \ar[r] \ar@{-}@(l,u)[dl]
        &\stackrel{u_3}{\chainport} 
        &\ntaritytwo{\ \ \mathtt{c:Client}\ \ }{r}{l}&
        \stackrel{u_4}{\chainport}
        \\
        \stackrel{u_2}{\chainport} 
        & \ntaritytwo{\flightobj{f_2}}{r}{l}
        & \stackrel{u_1}{\chainport}
      }
      \end{minipage}
%       &\stackrel{cf2}{\Longrightarrow}&
%       \xymatrix@C=.3cm@R=.3cm{
%         &\ntedge{\flightobj{f_1}} \ar[r] \ar@{-}@(l,u)[dl]
%         &\stackrel{u_3}{\chainport} 
%         &\ntaritytwo{\ \ \mathtt{c:Client}\ \ }{r}{l}&
%         \stackrel{u_5}{\chainport}
%         \\
%         \stackrel{u_2}{\chainport} 
%         & \ntaritytwo{\flightobj{f_2}}{r}{l}
%         & \stackrel{u_1}{\chainport}
%         &\ntaritytwo{\ \ \mathtt{c:Client}\ \ }{r}{l}&
%         \stackrel{u_4}{\chainport}
%       }
    \end{array}
    \]
\end{myex}
Observe that, unlike in the application of ADR productions, the
identity of edge $\mathtt{c}$ is preserved when applying the
reconfiguration rule $\mathtt{cf}$.
Also, for simplicity in Example\ref{ex:reconfiguration} we
take $\mathtt{c}$ to be just  a single edge, but the effect
of $\mathtt{cf}$ would be the same if instead of edge $\mathtt{c}$
we had a complex graph of type $\mathtt{Client}$: the whole
graph would have been moved from node $u_1$ to node $u_4$.

A result in~\cite{bllmt08} shows that the simple condition that 
an ADR rewriting system where all reconfiguration rules of
form~\eqref{eq:reconfiguration} have $t$ and $t'$ of the same sort
guarantees that the architectural style is preserved when
the system evolves.

In this paper we exploit the algebraic presentation of ADR production
and reconfiguration mechanisms and combine them together with a tracking
mechanism that is used to recover possible run-time misbehaviour. 
%%%%%%%%%%%%%%%%%%%%%%%%%%%%%%%%%%%%%%%%%%%%%%%%%%%

\defref{def:trackprods} below formalises our tracking 
mechanism using some trees to record graphs' evolution
due to productions and reconfigurations respectively.
We introduce some technical machinery first.

We consider forests of trees with vertexes drawn from a set
$\tnodeset$ (hereafter we will call the nodes of the trees vertices's in
order to distinguish them from the graph's nodes); if $f: X \to Y$ is
a partial map then we write $f(x)\undef$ when $f$ is undefined on $x$
and we let $\dom f = X \setminus \set{x \in X \st f(x)\undef}$.
Hereafter, we fix a finite set of productions $\pset$ to denote all the
productions of the system.
A \emph{tracking Environment} $\treeEnv$ is pair of two injective
finite partial maps
\[
\tenv 1: \tnodeset \to \edgeset \times \nodeset^*,
\qquad \text{and} \qquad
\tenv 2: \tnodeset \to \pset,
\]
and we use $\envempty$ to denote the empty environment (that is the
environment undefined on all $\lvar n \in \tnodeset$).

Basically, given a forest $T$, we use an environment $\treeEnv$ (such
that $\dom \treeEnv$ is the set of vertices's of $T$) so to decorate 
each vertex of $T$ with two attributes:
\begin{itemize}
\item $\treeEnv^{(1)}(\lvar n)$ assigns an edge with its list of
  nodes to the vertices's of $T$, and
\item $\treeEnv^{(2)}(\lvar n)$ assigns a production to the vertex
  $\lvar n$ in $T$.
\end{itemize}
It is convenient to write $\treeEnvNot{\lvar n}{e(\tilde x)} p$ when
$\tenv 1(\lvar n) = e(\tilde x)$ and $\tenv 2(\lvar n) = p$.
Also, in the following we use a notation inspired by object-oriented
programming to manipulate trees; more precisely, we consider trees $T$
(and their nodes $\lvar n$) as objects and write $T.\addTree(\lvar n,
T_1,\ldots,T_k)$ to add the trees $T_h$ as sub-trees of $T$ by rooting
them at the vertex $\lvar n$ in $T$; that is, the resulting tree will be
$T$ where vertex $\lvar n$ has the root of $T_1,\ldots,T_k$ as new
children.
Also, we let $\deg n$ to be the degree of a vertex $\lvar n$, 
$\child j {\lvar n}$ to be its $j$-th child, and (abusing notation)
we allow ourselves to identify trees consisting only of a root with 
the root vertex.
\begin{mydef}[Tracking productions]
  \label{def:trackprods}
  Let $G_0,\dots ,G_m$ be a sequence of graphs s.t. for each $0 \leq j
  < m$, $G_{j+1}$ is obtained from $G_j$ by applying a production
  $p_j\in \pset$ with morphisms $\sigma'_j : L_j \to G_j$ and
  $\sigma_j: R_j \to G_{j+1}$ where $L_j$ and $R_j$ are the LHS and
  RHS of $p_j$, respectively.

  We associate to each $G_j$ a \emph{tracking forest} $T_j$
  and a \emph{tracking environment} $\treeEnv_j$ as follows:
  \begin{itemize}
  \item let $r$ be the number of edges in $G_0$, forest $T_0 = \lvar
    n_1, \dots , \lvar n_r$ consists of $r$ single-vertex trees with
    roots $\lvar n_1, \dots , \lvar n_r$ taken pairwise distinct.
    Environment $\treeEnv_0$ is defined as the map that takes the
    $m$-th vertex in the forest $T_0$ to the $m$-th edge of $G_0$;
    formally, 
    \[
    \treeEnv_0 [\lvar n_m \mapsto e_m(\tilde x_m) \cdot
    \undef ] \quad \text{for} \quad 1 \leq m \leq r
    \]
    where $e_m = G_0[m]$ and $\tilde x_m$
    are the nodes in $G_0$ that $e_m$ is attached to;
  \item Let $k_j$ be the number of edges in the RHS of $p_j$ (that is,
    $k_j$ is the cardinality of $E_{R_j}$), and let $\lvar n$ be the inverse image of $\sigma'_j(e_j(\tilde x_j))$ through $\treeEnv^{(1)}_j$ s.t.
    \hidden{
      \[
      \begin{array}{c@{\qquad}c}
        \lvar n \text{ be } \begin{cases}
          \lvar n', & \text{if the inverse
            image of } \sigma'_j(e_j(\tilde x_j)) \text{ through } 
          \treeEnv^{(1)}_j
          \text{ is } \{\lvar n'\}
          \\
          \text{a fresh vertex} & otherwise
        \end{cases}
      \end{array}\]
    }
    $
    \treeEnv^{(1)}_j(\lvar n) = \sigma'_j(e_j(\tilde x_j))
    $
    and, for $1 \leq l \leq k_j$, let $T'_l$ be a tree made of just a
    fresh vertex node, then
    \[
    \begin{array}{r@{\hspace{.1cm}}c@{\hspace{.1cm}}l}
      T_{j+1} & = &
      T_j.\addTree(\lvar n, T'_1,\ldots,T'_{k_j}) %&
    \end{array}
    \]
    % where the inverse image of $\sigma'_j(e_j(\tilde x_j))$ through
    % $\treeEnv^{(1)}_j$ is the singleton $\{\lvar n\}$.
    Environment $\treeEnv_{j+1}$ is obtained by updating
    $\treeEnv_j$ in the following way:
    \[
    \begin{array}{r@{\hspace{.1cm}}ll}
      \treeEnv_{j+1} & = &
      \begin{cases}
        \treeEnv_j [\lvar n \mapsto \sigma'_j(e_j(\tilde x_j)) \cdot
        p_j , \ T'_l \mapsto \sigma_j(R_j[l]) \cdot \undef \st
        l = 1, \ldots, k] & \text{if} \ E_{R_j} \neq \emptyset
        \\
        \treeEnv_j [\lvar n \mapsto \sigma'_j(e_j(\tilde x_j)) \cdot
        p_j , \ T'_1 \mapsto \undef \cdot \undef] &
        \text{if} \ E_{R_j} = \emptyset
      \end{cases}
    \end{array}
    \]
  \end{itemize}
\end{mydef}
Despite some technical intricacy, Definition~\ref{def:trackprods} 
is conceptually simple.
Basically, we add to $T_j$ as many fresh vertexes as the edges in 
the RHS of the production $p_j$; such vertexes become the children 
of the vertex $\lvar n$ in $T_j$ associated with the LHS 
$\sigma_j'(e_j)$.
\hidden{
  \begin{itemize}
  \item either the node $\lvar n$ in $T_j$ associated with the LHS
    $\sigma_j'(e_j)$, had such a node been introduced by a previous
    production
  \item or of a fresh node that becomes the child of the root of
    $T_{j+1}$.
  \end{itemize}
}
Accordingly, the environment $\treeEnv_{j+1}$ updates $\treeEnv_j$ recording
edges and productions associated to $\lvar n$ and the 
fresh roots of $T'_l$.
Observe that each forest $T_j$ has $r$ trees, with $r$ the number of
edges of $G_0$.
Indeed, the evolution of $G_0$ involves only the replacement of such
edges (and those produced by such replacements).
Therefore, we can record the application of $p_j$ to $G_j$ in a node
of one of the trees representing the evolution of one of the initial
edges of $G_0$.

\begin{myex}\label{ex:dpTracking}
  Consider the production $\browseflightsprod$ from 
  Example~\ref{ex:reconfiguration}.
  For presentation purposes we use $p_j(e)$ to indicate the $j$-th 
  	application of the production $\browseflightsprod$ on an edge $e$.
  	
  {\small \[
    \begin{array}{c@{\hspace{.5cm}}c@{\hspace{.5cm}}l@{\hspace{.5cm}}c@{\hspace{.5cm}}l}
      \xymatrix@C=.3cm@R=.3cm{
        G_0:&
        \stackrel{u_2}{\chainport}
        &\ntedge{\flightobj{f}} \ar[r] \ar@{-}[l]
        &\stackrel{u_1}{\chainport}
      }
      &&
      \xymatrix@C=.3cm@R=.3cm{
        T_0:
        && \ \ \ \stackrel{x}{\chainport}
      }
      &&
      \begin{array}{rl}
        \treeEnv_O :& \lvar{x} \mapsto [f(u_1,u_2), \undef] 
      \end{array}
      \\
      \\
      \Downarrow p_0(f)
      \\
      \xymatrix@C=.3cm@R=.3cm{
        G_1:&        
        & \ntedge{\flightobj{f_1}} \ar[r] \ar@{-}@(l,u)[dl]
        &\stackrel{u_3}{\chainport}
        \\
        &\stackrel{u_2}{\chainport}
        & \ntaritytwo{\flightobj{f_2}}{r}{l}
        &\stackrel{u_1}{\chainport}
      }
      &&
      \xymatrix@C=.3cm@R=.3cm{
        T_1 :
        &&\stackrel{x}{\chainport}\ar@{-}[dl]\ar@{-}[dr]
        \\
        &\stackrel{x_1}{\chainport}&&\stackrel{x_2}{\chainport}
      }
      &&
      \begin{array}{rl}
        \treeEnv_1 :& \lvar{x} \ \mapsto [f(u_1,u_2), p_0]
        \\ & \lvar{x_1} \mapsto [f_1(u_3,u_2), \undef]
        \\ & \lvar{x_2} \mapsto [f_2(u_1,u_2), \undef]
      \end{array}
      \\
      \\
      \Downarrow p_1(f_2)
      \\
      \xymatrix@C=.3cm@R=.3cm{
        G_2:&        
        & \ntedge{\flightobj{f_1}} \ar[r] \ar@{-}@(l,u)[dl]
        &\stackrel{u_3}{\chainport}
        \\
        &\stackrel{u_2}{\chainport}
        & \ntaritytwo{\flightobj{f_3}}{r}{l}
        &\stackrel{u_4}{\chainport}
        \\
        && \ntedge{\flightobj{f_4}} \ar[r] \ar@{-}@(l,d)[ul]
        &\stackrel{u_1}{\chainport}
      }
      &&
      \xymatrix@C=.3cm@R=.3cm{
        T_2 :
        &&\stackrel{x}{\chainport}\ar@{-}[dl]\ar@{-}[dr]
        \\
        &\stackrel{x_1}{\chainport}
        &&\stackrel{x_2}{\chainport}\ar@{-}[dl]\ar@{-}[dr]
        \\
        &&\stackrel{x_3}{\chainport}&&\stackrel{x_4}{\chainport}
      }
      &&
      \begin{array}{rl}
        \treeEnv_2 :& \lvar{x} \ \mapsto [f(u_1,u_2), p_0]
        \\ & \lvar{x_1} \mapsto [f_1(u_3,u_2), \undef]
        \\ & \lvar{x_2} \mapsto [f_2(u_1,u_2), p_1]
        \\ & \lvar{x_3} \mapsto [f_3(u_4,u_2), \undef]
        \\ & \lvar{x_4} \mapsto [f_4(u_1,u_2), \undef]
      \end{array}
    \end{array}\]}
    Graph $G_0$ represents the initial graph, $T_0$ its corresponding 
    tree mapped to an initial environment $\treeEnv_0$ as defined in
    \defref{def:trackprods}.
    By applying $\browseflightsprod$ to $f$ we obtain $G_1$. Since
    the RHS of $\browseflightsprod$ generates two edges we have to add
    two nodes in $T_0$ as children of the node corresponding to $f$ to
    obtain $T_1$. Finally, we update the map of $\lvar x$ in 
    $\treeEnv_0$ to add the production applied to $f$ and add 
    two new mappings for the fresh nodes of $T_1$ to get $\treeEnv_1$.
    We now repeat this procedure for every application of a production.
\end{myex}

Hereafter, we will show the environment $\treeEnv$ within the tree $T$ 
as shown in the example below.
\begin{myex}\label{ex:treeandenv}
  This example shows a simplified way of representing the tree $T$ 
  together with the environment $\treeEnv$. We borrow the tree $T_1$ and 
  the environment $\treeEnv_1$ from Example~\ref{ex:dpTracking} to show 
  this visual simplification.
  \bigskip
  \center
  \begin{tikzpicture}%[scale=2]
    \node {[$f(u_1,u_2)$, $p_0$]} [grow'=down] 
    child {node [shift={(2,0)}] 
      {[$f2(u_1,u_2)$, $p_1$]}
      child {node [shift={(1.5,0)}] {[$f_4(u_1,u_2)$, $\undef$]}} child
      {node [shift={(-1.5,0)}] {[$f_3(u_4,u_2)$, $\undef$]}} } child
    {node [shift={(-2,0)}] {[$f_1(u_3,u_2)$, $\undef$]}};
  \end{tikzpicture}
  % \[
  %	\includegraphics[width=15cm]{./visio/example9.png}
  % \]
\end{myex}

%%% Local Variables: 
%%% mode: latex
%%% TeX-master: "main"
%%% End: 

%% file: rewrite.tex
Following the recording mechanism in~\defref{def:trackprods} 
we define a new rewriting mechanism for ADR reconfigurations.
Our approach hinges on tracking trees similar to those in
Definition~\ref{def:trackprods}.
The new rewriting approach will also allow us to keep track of changes
due to reconfigurations.

Given a term $t \in \talg$ and a tracking tree $T$, the bow tie relation
$\ptm t T$ holds iff $t$ and $T$ are isomorphic ``up to the leaves''
of $t$; more precisely, the tree obtained by considering just the
internal nodes of $t$ is isomorphic to $T$.
This is formalised in Definition~\ref{def:termmatch}
\begin{mydef}[Bow tie relation]
\label{def:termmatch}
The relation $\ptm{}{}$ in defined as
\[
\begin{array}{rll}
  \ptm t T & \iff &
  t \in \mathcal X
  \quad \text{or} \quad
  \treeEnv^{2}(\troot) = p \ \land\ t = p(t_1, \ldots, t_k) \ \land\ 
  \displaystyle{\bigwedge_{j = 1,\dots, k}} \ptm{t_j} {\child j \troot}
\end{array}
\]
where $\troot$ is the root of $T$.
\end{mydef}
Given a reconfiguration rule $\rho : t \to t'$, relation $\ptm \_ \_$
given in Definition~\ref{def:termmatch} allows us to identify which 
parts of a graph match the LHS of $\rho$ exploiting the 
correspondence between tracking trees and graphs.
A sub-tree $T'$ of  $T$ \emph{matches} $t$ iff $\ptm t {T'}$.
Assuming $\ptm t {T'}$ holds then given a variable $x$ occurring in
$t$, \defref{def:varmatch} returns the sub-tree of $T'$ (say  $T_x$)
corresponding to $x$. This is obtained by applying $\getvartree{t}{T'}{x}$
and is formalised below.

\begin{mydef}[Tree of a variable]
\label{def:varmatch}
Let $x$ be a variable occurring in $t$, $\troot$ be the root of $T$,
$1 \leq j \leq k$, and $x \in t_j$ denote that $t_j$ contains variable
$x$ then
\[
\begin{array}{rll}
  \getvartree{t}{T}{x} = &
  \begin{cases}
    T & \text{if} \quad t=x
    \\
    \getvartree{t_j}{\child j \troot}{x} & \text{if} \quad
    \treeEnv^{(2)}(\troot) = p \ \land\ t = p(t_1, \ldots, t_k) \
    \land\ x \in t_j 
    %\ \text{for}\ 1 \leq j \leq k 
  \end{cases}
\end{array}
\]
returns the subtree of $T$ corresponding to $x$ if $T$ can mach a path
from the root of $t$ to $x$.
\end{mydef}

\begin{figure}
\centering
\begin{tikzpicture}[->,>=stealth',node distance=1.2cm and 3cm,%shorten >=3pt,
  thick,main node/.style={ellipse,fill=blue!20,draw}, scale=1, every node/.style={transform shape}]
  \node[main node] (R) {$R$};
  \node[main node] (R') [below= of R] {$R'$};
  \node[main node] (L) [left= of R] {$L$};
  \node[main node] (L') [below= of L] {$L' (\hat{n}_1, \dots,\hat{n}_k)$};
  \node[main node] (E) [right= of R] {$E_j (\bar{n}_{j,1},\dots , \bar{n}_{j,l_j})$};
  \node[main node] (G) [below= of E] {$G_j (\breve{n}_{1_j},\dots , \breve{n}_{l_j})$};
  % \path%[every node/.style={font=\sffamily\huge}]
  \path  (L') edge [morphism] node [left] {$\iota^{-1}_L$} (L);
  \path  (R) edge [morphism] node [left] {$\iota_R$} (R');
  \path  (L) edge [map] node [above] {$i$} (R);
  \path (L') edge [bend right,morphism] node [below] {$\delta$} (R');
  \path  (G) edge [bend left,morphism] node [below] {$\sigma$} (R');
  \path  (G) edge [-,dashed] node [below] {} (E);
  \path  (E) edge [hookstyle] node [below] {} (R);
\end{tikzpicture}
\caption{Term to graph morphisms and maps (\cref~\defref{def:termToGraph})}
\label{fig:termtograph}
\end{figure}

\defref{def:termToGraph} builds a graph $\gamma (t)$ out of a term $t
\in \talg$. $\gamma$ inspects $t$ inductively and generates a graph
corresponding to the productions associated to $t$.  In the places of
the variables of $t$, $\gamma$ generates an edge of the appropriate
type.
More precisely $\gamma (t)$ returns a triplet of fresh edges, nodes,
and a mapping-function that relates variables of $t$ the fresh edges
generated.
For clarity, in \figref{fig:termtograph} we provide a visual view of the morphism
and mappings of \defref{def:termToGraph}.

\begin{mydef}[From Term to Graph]
  \label{def:applyRec}
  \label{def:termToGraph}
  Let $p = \anglebraces{ L, R, i:\nodes L \to \nodes R} \in \pset$ and
  $t \in \talg$.
  \[
  \begin{array}{c@{\qquad}c}
    \gamma (t) = 
    \begin{cases}
      (e,\ [n_1 , \dots , n_h],\  \eta : t \mapsto e)
      & \text{if } \ t \in \mathcal X, 
      \\
      ((G_1 \cup \dots \cup G_r)\sigma, \ \inter, \ \eta_1;\sigma \cup \cdots \cup \eta_r;\sigma)
      & \text{if } \ t = p(t_1 , \dots , t_r) , 
      \\ & \gamma(t_j)=(G_j , \inter_j , \eta_j) \text{ for } 1 \leq j \leq  r
    \end{cases}
  \end{array}
  \]
  where in the first clause,
  \begin{itemize}
  \item $e$ is a fresh edge of the type corresponding to $t$,
  \item $[n_1 , \dots , n_h]$ are its ($e$'s) fresh pairwise distinct
    nodes and
  \item $\eta$ is the mapping of the variable $t$ to the fresh edge $e$
  \end{itemize}
  and in the second clause,
  \begin{itemize}
  \item $\delta_j = [\breve{n}_{1_j},\dots , \breve{n}_{l_j}]$ for
    each $1 \leq j \leq r$ and
  \item if $\iota_L : L \to L'$ and $\iota_R : R \to R'$ are the
    isomorphisms from $L$ and $R$ to two fresh graphs $L'$ and $R'$
    (s.t. all their nodes and edges are fresh) respectively then
    \begin{itemize}
    \item $\delta = \iota_R(i(\iota_L^{-1}([\hat{n}_1, \dots,
      \hat{n}_k])))$ where $[\hat{n}_1 , \dots , \hat{n}_k]$ are the
      nodes of $L'$ and
    \item \ for $1 \leq m \leq l_j$, \ $\sigma: \breve{n}_{m} \mapsto \iota_R(\bar{n}_{j,m})$ where
      $[\bar{n}_{j,1},\dots , \bar{n}_{j,l_j}]$ are the nodes of the
      $j$-th edge of $R$.
    \end{itemize}
  \end{itemize}
\end{mydef}
Definition~\ref{def:termToGraph} builds a graph $\gamma(t)$ out of a term $t$.
Intuitively, $\gamma$ inspects $t$ ``bottom-up'' and it associates
disjoint designs (graphs $G_j$ with interfaces $\delta_j$) to each
sub-term of $t$ (note that fresh edges attached to fresh nodes are
associated to each variable of $t$); then $\gamma$ composes the
disjoint designs according to the production $p$ which is rendered by
replacing the nodes through the substitution $\sigma$
in Definition~\ref{def:termToGraph}.

Definition~\ref{def:applyRec} below establishes how to apply a 
reconfiguration rule to a graph.
\begin{mydef}[Applying reconfiguration rules]
  \label{def:applyRec}
  Fix a reconfiguration rule $\rho : t \to t'$ with $X$ being the set
  of variables of $t$, a graph $G$, and a tracking tree $T$ of $G$
  with the corresponding environment $\treeEnv$; let $T'$ be a
  sub-tree of $T$ such that $\ptm t T'$.
  For $x \in X$, let $T'_x = \getvartree{t}{T'}{x}$ be the
  sub-tree of $T'$ corresponding to $x$.
  An \emph{application of $\rho$ to $G$ wrt $T$} is a graph
  $G' = G \subs{G_L}{G_t}$ where
  \begin{itemize}
  \item $G_L = \displaystyle{\bigcup_{l=1}^m}
    \treeEnv^{(1)}(\lvar n_l)$ where $\lvar n_1, \dots , \lvar n_m$ are
    the leaves of $T'$, and
  \item $G_t = \gamma(t')\subs{\eta(x)}{G_x \st x \in X}$ where $G_x$
    is the sub-graph of $G$ corresponding to $x$ and it is defined as
    $G_x = \displaystyle{\bigcup_{l =1}^{h_j}} \treeEnv^{(1)}(\lvar
    n_l)$ with $\lvar n_{j,1}, \dots, \lvar n_{j,h_j}$ being the
    leaves of $T'_x$.
  \end{itemize}
  Finally, given the sub-trees $T'_x$ computed above, we replace the
  tree $T'$ in $T$ with a fresh sub-tree $T''$ corresponding to $t'$
  where we replace the vertexes corresponding to the variables of $t'$
  with the appropriate sub-trees $T'_x$.
  We then update the environment $\treeEnv$ so that it maps all the
  nodes of $T''$ up to the sub-trees $T'_x$ to the productions
  associated to them through $t'$.
\end{mydef}
We observe that, using Definition~\ref{def:termToGraph}, Definition~\ref{def:applyRec}
simply replaces the graph corresponding to $\gamma(t)$ with the graph
corresponding to $\gamma(t')$ where the edges corresponding to the
variables of $t'$ are replaced by the corresponding sub-graphs of $G$
identified through the proper morphisms.

\newcommand{\addc}{\mathtt{addC}}
\newcommand{\brf}{\mathtt{brF}}

\begin{myex}\label{ex:applyrec}
  Let us consider again the productions $\mathtt{cf}$ in
  Example~\ref{ex:reconfiguration}; for readability here we use $t$
  and $t'$ to refer to the LHS and RHS of $\mathtt{cf}$ respectively
  and we also abbreviate $\browseflightsprod$ with $\brf$:
  \[
  \begin{array}{c@{\hspace{2cm}}c@{\hspace{2cm}}c}
    \mathtt{cf}  \quad :  \quad \brf( x , \bookf(y,z) ) & \rightarrow & \brf( \bookf(x,z), y )
    \\
    t:
    \begin{minipage}[c]{.4\linewidth}
      \xymatrix@C=.3cm@R=.3cm{
        &&\stackrel{\brf}{\chainport}\ar@{-}[dl]\ar@{-}[dr]
        \\
        &\stackrel{x}{\chainport}
        &&\stackrel{\bookf}{\chainport}\ar@{-}[dl]\ar@{-}[dr]
        \\
        &&\stackrel{y}{\chainport} &&\stackrel{z}{\chainport}
      }
    \end{minipage}
    & \rightarrow &
    t':
    \begin{minipage}[c]{.4\linewidth}
      \xymatrix@C=.3cm@R=.3cm{
        &&\stackrel{\brf}{\chainport}\ar@{-}[dl]\ar@{-}[dr]
        \\
        &\stackrel{\bookf}{\chainport}\ar@{-}[dl]\ar@{-}[dr]
        &&\stackrel{y}{\chainport}
        \\
        \stackrel{x}{\chainport}&&\stackrel{z}{\chainport}
      }
    \end{minipage}
  \end{array}
  \]
  
  Fix an environment $\treeEnv$  and consider the graph $G$ below:
  \[
    G: 
    \begin{minipage}[c]{1.0\linewidth}
      \xymatrix@C=1cm@R=.3cm{ &&\ntedge{\flightobj{f_1}} \ar[r]
        \ar@{-}@(l,ur)[dl] &\stackrel{u_3}{\chainport}
        \\
        &\stackrel{u_2}{\chainport} &
        \ntaritytwo{\flightobj{f_3}}{r}{l} &
        \stackrel{u_1}{\chainport}
        &\ntaritytwo{\mathtt{c_1:Client}}{r}{l}&
        \stackrel{u_4}{\chainport}
        \\
        &&\ntedge{\flightobj{f_4}} \ar[r] \ar@{-}@(l,dr)[ul] &
        \stackrel{u_5}{\chainport} &\ntedge{\mathtt{c_2:Client}}
        \ar[ur] \ar@{-}[ul]& }
    \end{minipage}
  \]
  To apply $\mathtt{cf}$ to $G$ we have to identify a tree in the forest
  tracking $G$; this is done inspecting each tree of the forest with
  $\ptm\_\_$.
  Suppose that this yields the tree
  \[
  \xymatrix@C=.3cm@R=.3cm{
    T':
    &&\stackrel{[f(u_1,u_2):\flight, \brf]}{\chainport}\ar@{-}[dl]\ar@{-}[dr]
    \\
    &\stackrel{[f_1(u_3,u_2):\flight, \undef]}{\chainport}
    &&\stackrel{[f_2(u_1,u_2):\flight, \bookf]}{\chainport}\ar@{-}[dl]\ar@{-}[dr]
    \\
    &&\stackrel{[f_3(u_1,u_2):\flight, \undef]}{\chainport}
    &&\stackrel{[c(u_1,u_4):Client, \addc]}{\chainport}\ar@{-}[dl]\ar@{-}[dr]
    \\
    &&&\stackrel{[c_1(u_1,u_4):Client, \undef]}{\chainport}
    &&\stackrel{[c_2(u_1,u_4):Client, \undef]}{\chainport}
  }
  \]
  such that $\ptm{t}{T'}$, then
  \begin{enumerate}
  \item \label{varG} using $\getvartree{t}{T'}{}$ we can obtain the
    sub-tree of $T'$ corresponding to each variable of $t$ and get
    \[\begin{array}{c}
    \getvartree{t^x}{T'}{x} = \stackrel{[f_1(u_3,u_2):\flight, \undef]}{\chainport}
    \qquad\qquad    \qquad\qquad
    \getvartree{t^y}{T'}{y} = \stackrel{[f_3(u_1,u_2):\flight, \undef]}{\chainport}
    \\
    \getvartree{t^z}{T'}{z} = 
    \begin{minipage}[c]{1.0\linewidth}
      \xymatrix@C=.3cm@R=.3cm{ &&\stackrel{[c(u_1,u_4):Client,
          \addc]}{\chainport}\ar@{-}[dl]\ar@{-}[dr]
        \\
        &\stackrel{[c_1(u_1,u_4):Client, \undef]}{\chainport} &&\stackrel{[c_2(u_1,u_4):Client, \undef]}{\chainport} }
    \end{minipage}
    \end{array}\]
    By taking the union of all the edges mapped through $\treeEnv$
    to the leaves to the sub-trees corresponding to each variable we
    obtain sub-graphs of $G$ corresponding to each one.
    Observe that $z$ is not mapped to a single node $T'$.
    This is the reason we need to take the union of all the edges 
    corresponding to the leaves of $\getvartree{t^z}{T'}{z}$ in order
    to obtain its sub-graph.

  \item identify the sub-graph of $G$ corresponding to $T'$;
    similarly to step~\ref{varG} we take the union of all the edges mapped
    through $\treeEnv$ to the leaves of $T'$ to obtain:
    {\small
      \[
        G_{L}: 
        \begin{minipage}[c]{1.0\linewidth}
          \xymatrix@C=1cm@R=.3cm{ &&\ntedge{\flightobj{f_1}} \ar[r]
            \ar@{-}@(l,u)[dl] &\stackrel{u_3}{\chainport}
            &\ntedge{\mathtt{c_1:Client}} \ar@(r,ru)[dr] \ar@{-}[dl]
            \\
            &\stackrel{u_2}{\chainport} &
            \ntaritytwo{\flightobj{f_3}}{r}{l} &
            \stackrel{u_1}{\chainport}
            &\ntaritytwo{\mathtt{c_2:Client}}{r}{l}&
            \stackrel{u_4}{\chainport} }
        \end{minipage}
      \]
    }
  \item use $\gamma(t')$ to construct a graph corresponding to $t'$ 
    % in order to replace $G_L$ in $G$.
    {\small
      \[
        \gamma(t'): 
        \begin{minipage}[c]{1.0\linewidth}
          \xymatrix@C=1cm@R=.3cm{ &&\ntedge{\flightobj{e_1}} \ar[r]
            \ar@{-}@(l,u)[dl] &\stackrel{v_3}{\chainport}
            &\ntaritytwo{\mathtt{e_3:Client}}{r}{l}&
            \stackrel{v_4}{\chainport}
            \\
            &\stackrel{v_2}{\chainport} &
            \ntaritytwo{\flightobj{e_2}}{r}{l} &
            \stackrel{v_1}{\chainport} }
        \end{minipage}
      \]
    }
  \item $\gamma(t')$ represents the graph corresponding to $t'$
    where in the place of the variables it contains a dummy edge
    of the appropriate type. This is where step~\ref{varG}
    comes in place.
    We replace all the dummy edges in $\gamma(t')$
    with the graphs corresponding to the variables of the dummy edges
    to obtain $G_{t'}$.
    {\small
      \[ 
        G_{t'}: 
        \begin{minipage}[c]{1.0\linewidth}
          \xymatrix@C=1cm@R=.3cm{ &&\ntedge{\flightobj{f_1}} \ar[r]
            \ar@{-}@(l,u)[dl] &\stackrel{u_3}{\chainport}
            &\ntedge{\mathtt{c_1:Client}} \ar@(r,ru)[dr] \ar@{-}[l]
            \\
            &\stackrel{u_2}{\chainport} &
            \ntaritytwo{\flightobj{f_3}}{r}{l} &
            \stackrel{u_1}{\chainport}
            &\ntaritytwo{\mathtt{c_2:Client}}{r}{ul}&
            \stackrel{u_4}{\chainport} }
        \end{minipage}
      \]
    }
  \item the last step requires that we replace $G_L$ in $G$ with
    $G_{t'}$.
  \end{enumerate}
\end{myex}

%%% Local Variables: 
%%% mode: latex
%%% TeX-master: "main"
%%% End: 

%% file: methodology.tex
In~\cite{our:ice2012} we gave a basic methodology for recovering a
system to a valid state when a run-time configuration compromises
its architectural style. The main objective of our approach 
in~\cite{our:ice2012} can be observed in Example~\ref{ex:style} below.

\begin{myex}[~\cite{our:ice2012}]\label{ex:style}
  Consider the run-time reconfiguration
  \[\begin{array}{c@{\hspace{2cm}}c}
    \def\g#1{\save
      [].[dr]!C="g#1"\frm{}\restore}%
    \xymatrix@C=.5cm@R=.3cm{
      \ntarityone{S}{r} 
      & \stackrel{u}{\chainport}
      & \tarityone{C}{l}
      &&&&&
      % g2r1
      \ntarityone{F}{r}
      & \stackrel{u}{\chainport}
      & \tarityone{C}{l}
      \ar @{.>} "1,4" ;"1,7" ^-{badServer()}
    }
  \end{array}\]
  where $S$ changes as illustrated to model a failure $F$.
  By imposing an invariant that states that every client has to be
  connected to a non-failed server, the invalid configuration
  can be identified and recovered.
\end{myex}

The fact that we can now record the evolution of our graphs gives us
many advantages.
The monitoring mechanism introduced in~\ref{sec:tracking} and the 
way we apply reconfigurations now gives us the potential to 
identify which part of the graph (say $G$) has been re-written.
Using the tree corresponding to the affected sub-graph (say $G_R$)
one can identify the last production applied by observing the parent
node of the leaves corresponding to $G_R$.

By using the monitoring system one can now spot where a violation
may have occurred and attempt to fix such violations without 
considering or parsing the whole graph in every possible way.
%%%%%%%%%%%%%%%%%%%%%%%%%%%%%%%%%%%%%%%%%%%%%%%%%%%%%%%%%%%%%%%%%%%%%%%%%%%%%%%%%%%%%%%

Our methodology consists of the steps $1 \div 5$ below \footnote{
  Recall that the designer has to specify productions and the
  architectural invariant $\inv$ so to establish the architectural
  style of interest (as done in Example~\ref{ex:style}).}.
   Steps $\ref{mth:check} \div \ref{mth:pre}$ are inherited 
from~\cite{our:ice2012}.
\begin{enumerate}
\item \label{mth:parse} The architecture (say $G$) corresponding to
  the configuration of the current system is computed using the
  tracking tree $T$ and the corresponding environment $\treeEnv$ of
  the system. 
  % As also explained in Section~\ref{ssec:applyrewrite} we
  % take the union of all the edges mapped through the environment
  % $\treeEnv$ to the leaves of the sub-tree $T$ to obtain $G$.
\item \label{mth:mark} Identify the sub-tree $T_R$ of $T$ ($\ptm{t'}{T_R}$)
  that corresponds to the RHS term $t'$ of the reconfiguration applied.  
\item \label{mth:check} Check whether $G$ satisfies $\inv$ 
  ($G \models \inv$).
  \begin{enumerate}
  \item If $G \models \inv$ then the style is not violated.
  \item If $G \nvDash \inv$ then go to step~\ref{mth:pre}
  \end{enumerate}
\item \label{mth:pre} For each production $p$, compute the weakest 
  pre-condition $\psi$ with respect to $\inv$.
  \begin{enumerate}
  \item \label{mth:last} Select a production $p : L \to R$ and
    let $\sigma: L \to G$ be the morphism from $L$ to $G$ 
    such that $\gnoL \models \psi$ (if
    any); apply $p$ to $G$ to determine the reconfiguration
    needed for the system to reach a valid state.
  \item \label{mth:wpcofwpc}\label{mth:reapply}
    If the designer considers not satisfactory
    the reconfigured system obtained in the previous stage or if there
    is no production $p$ such that $\gnoL \models \psi$, then the
    designer may either, repeat step~\ref{mth:last} by replacing $G$
    with $\gnoL$ and $\inv$ with $\psi$ or, try
    step~\ref{mth:parseagain}.
  \end{enumerate}
\item \label{mth:parseagain} Given $T_R$ computed in
  step~\ref{mth:mark} the designer can select a 2-tier subtree $T'_R$
  of $T_R$ that contains a parent vertex and all its leaf children
  correspond to replaceable edges through the environment
  $\treeEnv$.
  Using $T'_R$ in \defref{def:parse} we parse the graph as defined
  below and repeat step~\ref{mth:pre}.
\end{enumerate}
\begin{mydef}[Parsing]
  \label{def:parse}
  Given a graph $G$, a corresponding 2-tier sub-tree $T$ with root
  $\troot$ and an environment $\treeEnv$.  $p: L \stackrel{i}{\to}
  R$ is the production returned from $\tenv{2}(\troot)$.
  Given an instance $L'$ of $L$ through the isomorphism $\iota : L \to
  L'$ and let $\sigma' :R \to G$ a graph $G'= G
  \setminus (E_{G} \cup \sigma'(\inodes R)) \cup L''$ ($\inodes{R}$
  refer to the internal nodes of $R$) is the graph obtained by parsing
  $G$ with $p$ under the morphism $\sigma'$ iff $L'' =
  L'\subs{\iota(l)}{\sigma'(i(l)) \ \st\ l \in Im(i)}$ and there exist
  no edge in $G$ that is non-replaceable.
\end{mydef}

Identifying the architecture of a system is non-deterministic in the
original specification of ADR (\cref~\cite{bllmt08}). Using our approach
we can identify and retrieve the architecture of the system with all the
information of how it has been adapted so far using our monitoring
mechanism. This mechanism is used in step~\ref{mth:parse} for
retrieving graph $G$ and in step~\ref{mth:mark} for identifying the
subtree of the tracking tree that has been reconfigured.
In step~\ref{mth:check}, we assume that an underlying monitoring
mechanism uses the $\models$ relation of our logic to determine if the
graph $G$ identified in step~\ref{mth:parse} violates the invariant.
In such case, step~\ref{mth:pre} uses the weakest pre-condition
algorithm \cite{our:ice2012} on each production to compute
their weakest pre-conditions (this step does not need to be
re-iterated at each reconfiguration but instead all the weakest
pre-conditions can be computed offline and reused accordingly).
In step~\ref{mth:last}, if the graph representing the violated system
satisfies one of the computed weakest pre-conditions then the
corresponding production is a candidate to re-establish the
architectural style and trigger the appropriate reconfigurations on
the invalid system.
In step~\ref{mth:wpcofwpc} the designer has to decide whether
to stop or continue the process.
In the latter case, the idea is to repeat steps~\ref{mth:pre} and
\ref{mth:last} replacing $G$ with $\gnoL$ and $\inv$ with $\psi$ so to
 compute the weakest pre-condition of the weakest
pre-condition computed in the previous iteration.
This, allows us to exploit every possible sequence of productions 
that can be applied in order to enforce the architectural style.
Note that the morphism that invalidates $G \models \inv$ indicates
which part of the system has to be rewritten, while the production $p$
suggests plausible reconfigurations.
If the above steps do not prove to be satisfactory then in
step~\ref{mth:parseagain} the designer uses the tree computed in
step~\ref{mth:mark} and selects which branch of the tree can be
abstracted using the parsing mechanism in \defref{def:parse}.
After parsing the graph we repeat steps \ref{mth:pre} $\div$ 
\ref{mth:parseagain} until either, a possible sequence of productions
is identified that can potentially fix the architectural style, or,
the designer  decides to stop the process.

%%% Local Variables: 
%%% mode: latex
%%% TeX-master: "main"
%%% End: 

%% file: future.tex
In this paper we defined a framework that allows us to exploit the
``hierarchical nature'' of ADR graphs.
In particular, we used the functional reading of ADR productions as
well as its reconfigurations that preserve the identity of
components throughout the rewriting.
Our framework permits to tackle certain architectural aspects of the
design and allows the designer to identify and address problems at the
architectural level.

Also, we refine the methodology proposed in~\cite{our:ice2012} to
automatically compute possible reconfigurations that recover from
architectural style violations.
Our refinement iteratively computes the weakest pre-condition to find
a possible sequence of reconfigurations (if any) that re-establishes
the architectural style of the system.

In this  paper we focused on the development of the technical presentation
of our framework.
We proposed a monitoring mechanism through which the
evolution of a computation is recorded and maintained in a tree-like
structure reflecting the hierarchical nature of ADR graphs.
We then exploit this mechanism to formally define more efficient
parsing algorithms as well as more efficient ways of applying
reconfigurations.
Interestingly, this approach brings forth the definition of a new 
rewriting mechanism for ADR reconfiguration rules.
More precisely, instead of parsing an ADR graph searching for a
sub-graph matching the RHS of a reconfiguration rule, we propose
a rewriting mechanism that visits the trees describing the graph
evolution to find the match.
We argue that this is more efficient than parsing the graph at
the negligible cost of recording the evolution of the system
through the monitoring.
One could argue that the run-time monitoring of the system could be
inefficient.
In this respect, we note that a form of monitoring is necessary when
dealing with self-configuring or self-healing systems;
quoting~\cite{kc03} ``autonomic system might continually monitor its
own use''.
Since autonomic systems are the reference systems of this work, and,
as a matter of fact, a form of monitoring is indispensable in order to
identify run-time violation of the style, we argue that our approach
simply adds to the necessary monitoring activities the cost of
tracking the evolution of systems.

This paper completes and refines the framework initially proposed
in~\cite{our:ice2012}.
We contemplate a new research direction to devise autonomic systems
where component managers use architectural elements in the
reconfigurations they distill.
From this point of view, ADR is particularly suitable due to the fact
that it can represent not only architectural level aspects of systems,
but it can also be used to represent operational semantics by e.g.
encodings of process calculi or modelling languages~\cite{bgl10,bflmt11}.
This would immediately establish a connection between DbC approaches
of abstractions levels close to the implementation of systems (for
instance, see~\cite{bhty09,Liu02sc}) with the DbC approach for ADR
suggested in~\cite{our:ice2012}.

We conclude by commenting about the linearity conditions imposed on
the reconfigurations rules (c.f.~\eqref{eq:reconfiguration} and
\defref{def:applyRec}).
The linearity of the LHS of a reconfiguration rule can be relaxed
at the cost of making the semantic of the  matching more complex
since multiple occurrences of the same variable would account for
checking the existence of an isomorphism among different subgraphs.
Instead, the linearity condition on the RHS of a production can
be relaxed by simply using the counterpart semantic mechanism
described in~\cite{DBLP:journals/fuin/GadducciLV12} to keep track
of one copy of the variable.
Finally, we note that in~\cite{bllmt08}, reconfigurations rules of the
form $r(x) : p(y) \to q(x,y)$ are considered, where $x$ act as a
parameter of the rule that can be used in its LHS or RHS.
Such rules can be easily added to our framework using
\defref{def:applyRec} by mapping $\eta(x)$ to the fresh input graph.

%%% Local Variables: 
%%% mode: latex
%%% TeX-master: "main"
%%% End: 